# Flat metaform near-eye visor

CHUCHUAN HONG,[1] SHANE COLBURN[1], ARKA MAJUMDAR[1,2*]

[1]*Department of Electrical Engineering, University of Washington, Seattle, WA 98195*
[2]*Physics Department, University of Washington, Seattle, WA 98195*
*\*Corresponding author: arka@uw.edu*

**A near-eye visor is one of the most vital components in a head-mounted display. Currently, freeform optics and waveguides are used to design near-eye visors, but these structures are complex and their field of view is limited when the visor is placed near the eye. In this paper, we propose a flat, freeform near-eye visor which uses a sub-wavelength patterned metasurface reflector. The visor design imparts a spatial phase profile on a projected display pattern and can be implemented using a micron-scale thick metasurface. As the resulting metaform visor relies on diffraction, it can preserve a large field of view (77.3° both horizontally and vertically) when placed only 2.5 cm away from the eye. We simulate the metasurface visor to estimate the modulation transfer function, and find that the projected image quality is sufficiently high for human vision. While the design of the metasurface is initially performed via ray optics, using full-wave finite-difference time-domain simulation we validate a scaled version of our visor design.**

## 1. INTRODUCTION

In recent years, near-eye visors (NEV) have generated substantial interest among researchers in both academia and industry for their potential to be used in head-mounted displays (HMD), enabling a seamless augmented and virtual reality experience. In its simplest form, a NEV is an image magnification system that projects the information coming from a micro-display into a user's eyes. For a good user experience, HMD systems need to be compact and lightweight, while maintaining a large field-of-view (FOV). Most existing NEVs operate based on ray optics principles, i.e., reflection and refraction. This poses a stringent trade-off between the size of the NEV and FOV, i.e., bringing the NEV closer to the eyes decreases the FOV. As the visor comes closer to the eye, the light needs to be reflected at steeper angles to maintain the desired FOV. Reflection from a smooth surface cannot provide such arbitrary bending of light, which ultimately restricts the overall volume of HMD systems. Several designs have been reported in the literature to reach large FOV, including designs based on freeform optics [1-4], optical waveguides [5], reflective systems [6, 7], and retinal scanning technology [8]. For example, Y. Zhu et al. [6] designed an eight-mirror reversed telescope system to accomplish an ultra-thin near-eye device; D. Cheng et al. [9] combined geometrical waveguides technology with freeform optics technology for the design of an ultra-thin near-eye display; J. Yang et al. designed a see-through near-eye display using geometrical waveguides to accomplish a large FOV [5]; O. Cakmakci et al. [1] proposed a freeform single-element head-worn display using a 289 term Gaussian radial basis function for representing a freeform optical surface as both a magnifier and reflector. In all these designs, however, the FOV is still limited. In the single-element NED design, the full diagonal FOV is around $24^o$ [1]. The design from J. Yang et al. [5] with large FOV is limited to a horizontal FOV of only $30^o$ and vertical FOV of $60^o$.

In this paper, we propose a single-element metasurface-based NEV design, which relies on diffraction for its operation, and thus the light can bend at angles larger than what is possible using simple reflection. An optical metasurface is a quasiperiodic array of sub-wavelength optical antennas or scatterers which can modify an incident optical wavefront with sub-wavelength spatial resolution. Metasurfaces are similar to conventional diffractive optics, but due to the sub-wavelength nature and the resonant properties of their scatterers, metasurfaces can impart multi-level phase-shifts in the 0 to 2π range by only modifying the lateral geometry of the antennas. In conventional diffractive optics, such multi-level phase-shifts require elements with different thicknesses. This uniform thickness enables fabrication of metasurfaces using only a single lithography stage, whereas for conventional diffractive optics multi-stage lithography is necessary. Thus, metasurfaces enable flat and extremely thin (~1$\mu m$) optical elements and can be easily integrated into optical systems while maintaining ultra-compact size and weight. In recent years, several research groups have demonstrated various optical elements based on metasurfaces [10-14]. Along with rotationally symmetric structures, metasurfaces also show promise for building structures with arbitrary phase profiles, such as those for generating holograms and freeform optics [15, 16]. However, the primary application areas of these metasurfaces to date remain imaging, spectroscopy, and microscopy, and their use in compact NEV systems have not yet been explored.

Here, we propose a NEV design based on metasurface freeform optics, termed here as metaform optics, with sub-micron thickness [15]. Via numerical simulation, we find that the proposed metaform visor achieves high-quality images and a wider FOV than previously reported results: 77.3$^o$ in both the horizontal and vertical directions, when the visor is placed only 2.5 cm away from the eyes. The size of the visor is assumed to be

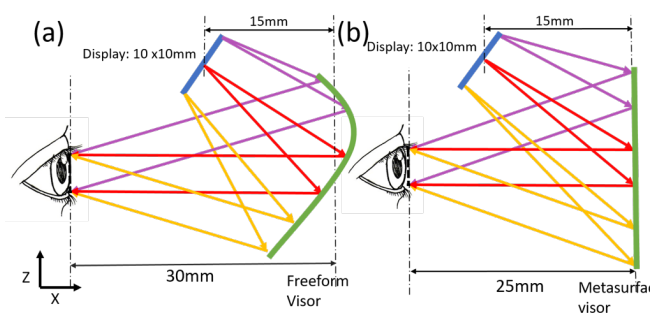

Fig. 1. Schematic of the proposed freeform NEV: (a) the XZ view of a free-form reflector: the eye and the display are shown enlarged for clarity; (b) a carefully designed metasurface visor can improve the FOV significantly, while bringing the visor closer to the eye.

4$cm$×4$cm$ to maintain a compact form-factor close to that of a pair of sunglasses. Our design achieves a modulation transfer function (MTF) exceeding 30% at 33 cycles/mm, with grid distortion less than 8.76%. These parameters are sufficient for human intelligibility [5]. The extreme thinness of the visor and its flat geometry facilitate integration with flexible substrates [17, 18]. With this ease of integration, one can envision building an adhesive and flexible sticker-like visor element which could be placed on any eye-wear to convert it to a NEV as seamlessly as placing a price tag on a pair of sunglasses. In conjunction with the NEV, a mini display placed on top would create a full HMD system.

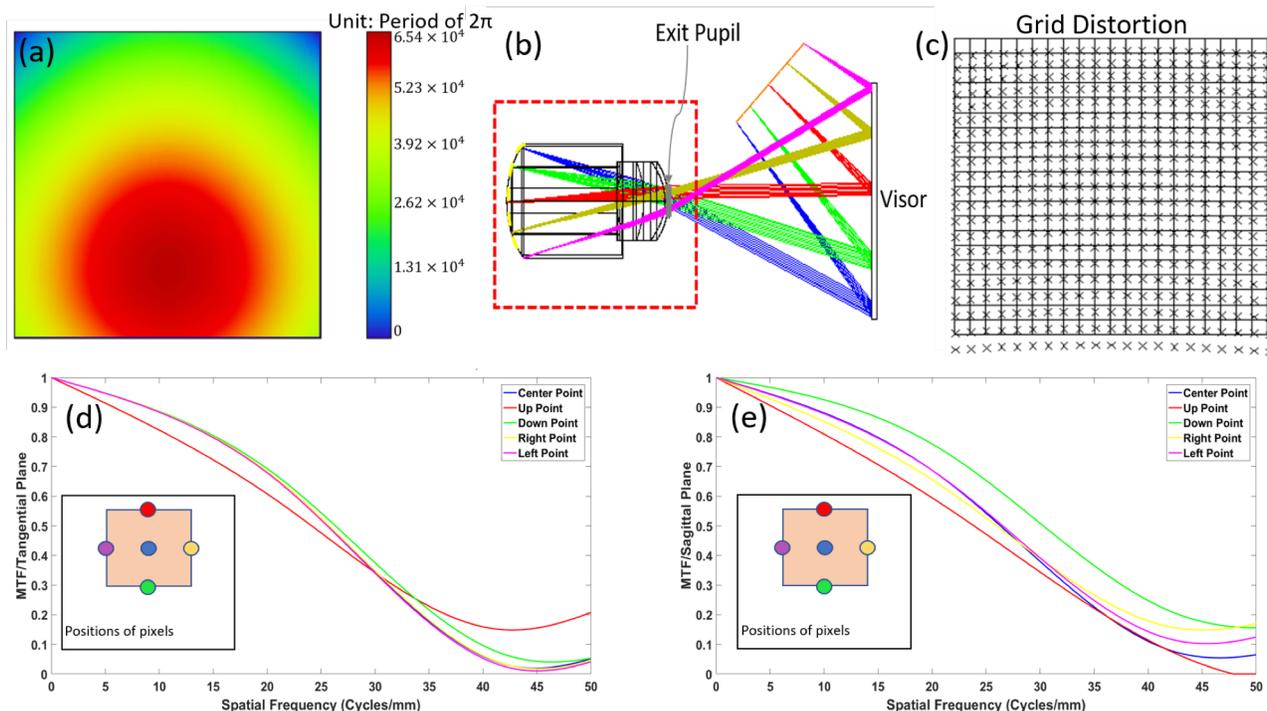

Fig. 2. Freeform near-eye visor based on a phase-mask: (a) Phase mask used in the Zemax, without phase wrapping. (b) Ray tracing simulation in ZEMAX: light rays from the display are reflected from the phase-mask and enter the eye-box (shown within the red dotted line). The visor and the exit pupil are labeled in the figure. On the very left end of each ray, the yellow arc-shaped dotted-line indicates where the retina is. (c) Grid Distortion of the NEV. Highest calculated distortion is 8.76%. (d) and (e) show the MTF of the NEV on tangential and sagittal plane, respectively. Distinct colors represent rays from different pixels, whose positions are shown in the inset of (c) and (d).

## 2. DESIGN OF METAFORM NEAR-EYE VISOR

First, we design a freeform optical element which can guide light from a HMD to the human eye (Fig. 1a). We use the widely accepted eye model proposed by Liou & Brennan at 1997 [19] for our simulations. For the best performance, we need to make sure that parallel rays of light enter the eye pupil (diameter assumed to be ~4mm). To obtain the freeform NEV design, we divide the display and the FOV into 10 segments. The NEV is designed so that the light from each segment of the display goes straight to the eye upon reflection from the visor. Thus, the NEV is designed as a collection of multiple small segments of plane mirrors with different orientations, and the resulting visor is essentially a reflective freeform surface. The number of segments is chosen to be 10, as further segmentation of the FOV does not appreciably change the shape of the surface. We find that as we bring the visor closer to the eyes, the incident light needs to be deflected at larger angles, and often the mathematical solution becomes unphysical; the shape of the visor will extend beyond the edge of the display, reaching its backside, which is a mathematically valid solution, but makes portions of the visor physically inaccessible by light. This limitation comes from the fact that the light guiding principle is based on reflection, and the incident light cannot be bent arbitrarily. Hence, we can only achieve a vertical FOV of 63° and horizontal FOV of 52° using the freeform visor relying on reflection. Moreover, the resulting NEV itself has a complicated shape (Fig. 1(a)) and significantly differs from that of an ordinary pair of sunglasses.

These problems can be avoided by using a metasurface-based NEV. The complicated shape and functional form of the NEV can be implemented using an ultrathin metasurface. To achieve the spatially varying angular deflection necessary to implement a metasurface-based visor, we use the generalized Snell's law in the presence of a periodic patterned interface [10]:

$$\sin(\theta_r) - \sin(\theta_i) = \frac{\lambda_0}{2\pi n_i} \frac{d\phi}{dx}$$

where, $\theta_r$ ($\theta_i$) is the reflection (incident) angle, $\lambda_0$ is the optical wavelength, $n_i$ is refractive index of the incident medium, and $\frac{d\phi}{dx}$ is the phase gradient along the interface. Note that in ordinary reflection, we assume a smooth interface, resulting in $\frac{d\phi}{dx} = 0$ and $\theta_r = \theta_i$. The reflected and incident angles are calculated from the surface normal. To calculate the deflection angle, the sign convention is opposite for incident and reflected light. Hence, if the reflected and incident rays are on the same side of the surface normal, they have opposite sign. By using a metasurface, we can arbitrarily engineer the phase gradient $\frac{d\phi}{dx}$, and thus can bend light in different directions. This directionality arises from a grating-like effect, and has been previously used in auto-stereoscopic, multi-view displays [20]. Based on this intuition, we first calculate the deflection angle required at different parts of the visor to faithfully project the display to the eye (Fig. 1b). We then calculate the phase-gradient to realize that deflection angle (i.e., $\theta_i + \theta_r$). During this calculation, the reflected light is bent anomalously to arbitrary directions to ensure the display image is accurately reproduced at the eye. For example, to have a large FOV in Fig. 2b, near the bottom of the metasurface visor the reflected and incident light need to be on the same side of the surface normal (shown in blue rays). Such a condition can be satisfied only by exploiting the metasurface's large phase gradient. Thus, we calculate the whole phase-mask as a collection of several segments of different phase-masks with varying phase-gradients.

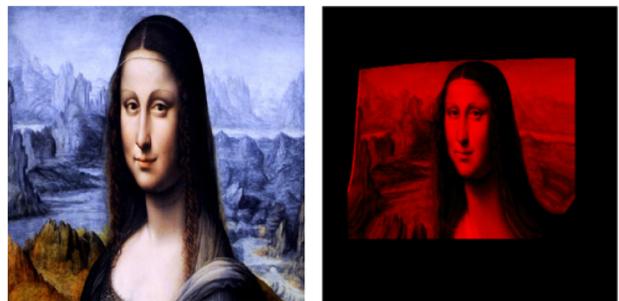

Fig. 3. Image Simulation of the Mona Lisa in the NEV using Zemax. The left figure is the original image projected in the display. The right figure is the simulated image as seen by the person using the NEV.

This phase-mask generally has an arbitrary form, and a closed form expression cannot be obtained.

We apply this methodology to a metaform NEV placed 2.5 cm away from the eye (Fig. 1b). The display is placed between the visor and the eye: 1cm away and 1.5 cm upwards from the visor with an angle of 45° with respect to the optical axis. From our simulations, we estimate the FOV to be 77.3° along both vertical and horizontal directions. The phase mask is calculated and shown in Fig.2a in units of $2\pi$.

## 3. PERFORMANCE OF THE VISOR

We first analyze the performance of the metaform NEV using ray optics. Here, we model the NEV as a continuous phase-mask and simulate the system using Zemax. Fig. 2b shows the ray tracing simulation setup. The observed image in the eye is a mirror-image of the display as expected. For these simulations, we did not wrap the phase modulo $2\pi$ when calculating the form of the element. Fig. 2c shows the grid distortion, with the largest distortion of 8.8% occurring at the lower corners. To estimate the quality of the optical image, we evaluate the modulation transfer function (MTF) of the NEV. Figs. 2d and 2e show the calculated MTFs for different points on the display. Fig. 2c is for the MTFs along the tangential plane while Fig. 2d is for those along the sagittal plane. In both MTF figures, the MTFs stay beyond 30% at 33 cycles/mm. They goes down to 10% at around 40 cycles/mm, which is sufficient for a visual system [5].

The primary achievement of the proposed NEV is its large FOV ( ~77° along both the vertical and horizontal directions), while keeping the distance of the NEV from the eye small. We compare our design with a previously reported freeform visor in Ref. [1]. Via Zemax simulation of their design, they had a full diagonal FOV of 24°, with around 3cm distance from eye to visor. Using their design, the FOV decreases further as the visor is brought closer to the eyes. The MTF and the grid distortion of our design is comparable to those of their design.

Additionally, we simulate an image of the Mona Lisa in Zemax to assess the performance of the NEV, as shown in Fig. 3. The projected image of the Mona Lisa is shown in Fig. 3a. The image reproduced on the retina after reflecting off the meta-form NEV and passing through the eye model is shown in Fig. 3b. Because the meta-form visor is designed for red wavelength, 633nm, the simulated image is in red color. The original image was square with each side of 1cm length. The image projected in the retina has a size of 7mm on each side. From grid distortion simulations, we find that the maximum distortion happens at the image corners. Such distortion, however, can be undone by applying an inverse operation to the image in advance, and projecting a filtered image. For example, we see the image output is warped and it could be easily revised by applying a simple 'de-warping' processing on the input image [21, 22]. Furthermore, the clarity of image is well-maintained, consistent with the previously calculated high-quality MTF results.

## 4. METAFORM VISOR SIMULATION

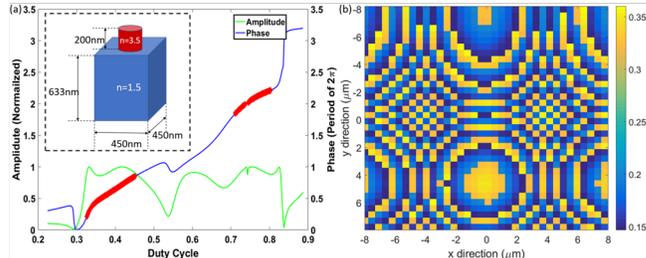

Fig. 4. Metasurface design: (a) result from the RCWA simulation using parameters described in the inset. Green line is the normalized amplitude response and blue line is the phase response in unit of $2\pi$. Duty cycle is defined as the ratio of the post diameter to the pixel size, or the periodicity. Inside red highlighted region, phase response is ranged from 0 to $2\pi$ range with high enough amplitude (larger than 0.8) simultaneously. For our simulations, we used only this range. (b) the distribution of pillar diameters on metasurface.

So far, we assumed the NEV to be a simple phase-mask. In practice, this phase-mask will be implemented using a metasurface. As metasurfaces are made of sub-wavelength scatterers, ray optics simulations alone cannot fully capture the underlying physics of the system. A complete full-wave simulation is warranted to establish the efficacy of the metasurface method for building the NEV. Unfortunately, a full-wave simulation of the actual NEV is impossible due to its large size. Hence, we scaled down the whole system by a factor of 2500 in all three dimensions, and simulated the imaging performance using commercially available Lumerical FDTD (finite-difference time-domain) software. In these simulations, the scatterers impart phase shifts in the range from 0 to $2\pi$, and the actual nano-photonic structure is simulated by considering the full vectorial nature of the electromagnetic field.

The metasurfaces are designed using cylindrical silicon pillars ($n \sim 3.5$) arranged in a periodic grid. By changing the pillar diameters, we can provide different phase-shifts. The phase-shifts and reflectivity as a function of the pillar diameter are first calculated using rigorous coupled-wave analysis (RCWA). In this simulation, we assume a periodic structure (Fig. 4a). Fig. 4a thus provides a map between the phase-shift and pillar diameter, based on which we can arbitrarily place different pillars in a periodic grid to mimic the desired phase function. Here, we implemented the phase profile obtained from the previous ray-optics simulation. We performed simulations at a wavelength of $\lambda = 633\,nm$, with a grid period of $450\,nm$, and a pillar height of 200 nm (Fig. 4a inset). The pillars are placed on a silica substrate ($n \sim 1.5$). As can be seen from Fig. 4a, the phase is changed in the range of $0 - 6.6\pi$, and there is significant amplitude modulation. To ensure that our down visor. The pillar diameter distribution over the whole surface is shown in Fig. 4(b).

For the image simulation using FDTD, we use a superposition of 100 Gaussian sources to mimic a two-dimensional display of size 10×10 [23]. We then create different shapes by appropriately setting the intensity of all the hundred sources. We ran simulations with 5 different shapes: a circle, rectangle, triangle, W (upside down), and U (upside down), as shown in Fig. 5. Fig. 5a shows the actual image, that we intend to project. However, as we have very low resolution in FDTD, the actual projected image is already distorted as shown in Fig. 5b. We used full-wave simulation to propagate the field from the display to the metaform visor, and from the visor to the opening of the eye-ball. This simulation truly models the electromagnetics of the nano-scale patterned posts can meet the requirements for both high reflected light intensity and generating phase shifts ranging from 0 to 2π simultaneously, we select diameters with both wide phase variation as well as high reflectivity (larger than 0.8). Based on the relationship between a specific diameter and the phase generated by the corresponding post, we mapped out the radii distribution for the scaled-metasurface, unlike our previous ray optical simulation, where we used an ideal phase mask. The effect of the eye's lens is simulated using an angular spectrum propagator which solves the Rayleigh-Sommerfeld diffraction integral [24]. Fig. 5d shows the images created by the metasurface implementing the scaled metaform visor as generated by FDTD simulation under red light illumination. We find that the projected images are faithfully reproduced in simulation; however, the granularity and distortion of the projected image are clearly visible. We believe that these imperfections primarily originate from the limited mesh size of the full-wave simulations. Additionally, we are further constrained by our computational resources in that the resolution of our display is limited to only 10×10 Gaussian sources. Hence, as shown in the second row of the Fig. 5, the actual input shapes on our display also exhibit non-uniform intensity profiles, and consequently the images on the retina inherit this non-uniformity. For example, in the case of the reverse U or W images in the second row, the intensities are concentrated at corners. We can also observe lower intensities on the horizontal lines in the reverse U and W images as well. Thus, for the U and W images in the fourth row in Fig. 5, horizontal lines also exhibit lower intensities than elsewhere, although the images are still clearly discernible.

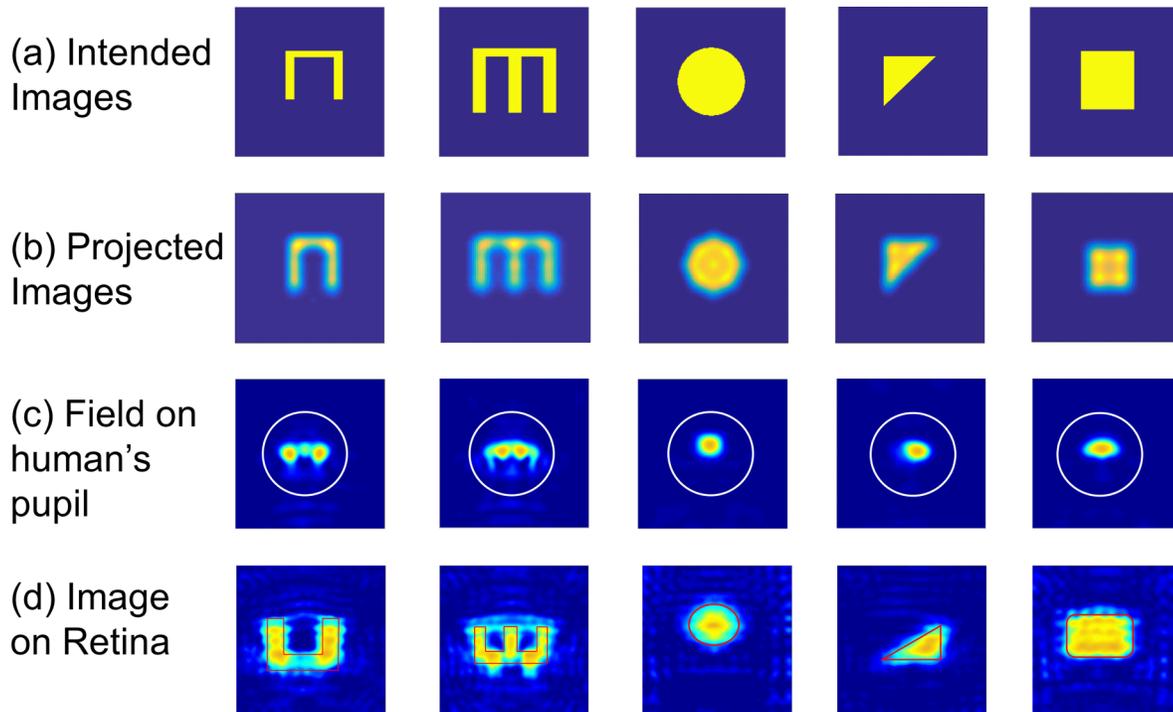

Fig. 5. Imaging simulation using Lumerical FDTD: (a) The first row shows the desired shapes. (b) Second row shows the images to be projected as modeled in the Lumerical FDTD. (c) The third row shows the electromagnetic field intensity at the plane of the human's pupil. (d) The last row shows the image in the retina, which is the image seen by a person. Five different shapes are simulated. For clarity, red lines highlight the shape we want to reproduce. It should be noted that in the third row, fields are all concentrated in the region of entrance pupil of eye, which is denoted as white circle.

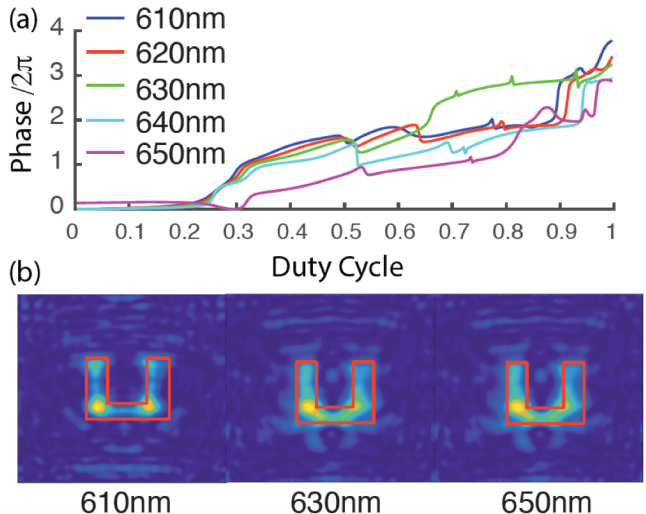

Fig. 6. Chromatic behavior: (a) the calculated phase shift using RCWA in the metasurface elements for five different wavelengths between 610nm and 650nm; (b) Simulated images via full-wave simulations for three representative wavelengths.

## 5. DISCUSSION

The proposed meta-form visor can overcome the FOV limitations of existing NEVs, especially when the NEV is placed close to the eyes. As the metasurface relies on diffraction, it can bend light at much steeper angles. Moreover, the metaform visor is flat and ultra-thin, and can be easily integrated with existing eye-wear. Thus, in using metaform visors, we can significantly reduce the volume of the whole NEV system. The possibility of transferring thin metasurfaces on a flexible substrate will allow for creating a flexible and adhesive sticker-like NEV which could be integrated with a conventional pair of glasses.

In the current design, the metasurface is made of silicon and provides high reflectivity because of its index contract, which is suitable for virtual reality technology. For augmented reality technology, we need to design a partially reflective "see-through" metasurface. This could be potentially realized by replacing silicon with a higher bandgap material (e.g., silicon nitride) to make the subwavelength posts more transparent [13]. Another serious limitation of the proposed metaform visor will be their strong chromatic aberrations, as are commonly observed in various metasurface optical elements. The chromatic aberrations in the metasurface originate from both the resonant nature of the metasurface elements, as well as the wrapping of the phase which produces discontinuities under wavelength deviation [25]. All the imaging simulations reported in Fig. 5 are performed under red light illumination (~633nm). When illuminated under green or blue light, no clear image is observed. However, the metasurface behaves reasonably well over a bandwidth of ~30nm about the designed red wavelength. We calculated the phase-shift using RCWA for different wavelengths near 633nm, and found that the phase does not change appreciably over ~30nm optical bandwidth. However, the phase becomes significantly different between 610nm and 650nm. We also performed imaging simulations to find that the image is discernible over ~40nm optical bandwidth. The chromatic aberrations can be corrected by using multi-wavelength metasurfaces, operating at red, green, and blue wavelengths. Several recent research studies have demonstrated operation of multi-wavelength metasurface optics [25, 26], and it is possible to extend such designs to the proposed metasurface NEVs as well. Using three stacked plasmonic metasurfaces, researchers also demonstrated operation of a metasurface lens at red, green and blue wavelengths [27]. Another approach could be to use an intelligent phase mask to extend the depth of focus to ensure that the point spread function of the metaform visor is the same for different colors [28]. For imaging, wavefront encoding has been used previously to perform such point spread function engineering [29]. Similar techniques could also be used for display technologies in the context of NEVs.

Our work for the first time explored the possibility of using emerging nanophotonic devices and metasurfaces to create compact near-eye visors. Starting from a geometric optics framework, we proposed a method to create the NEV, whose performance is validated by ray optics simulation. We also introduced a new method to perform imaging simulations using full-wave FDTD techniques and analyzed the performance of a near-eye visor metasurface. The large FOV while maintaining a compact form-factor indicates that nanophotonic devices can significantly benefit augmented and virtual reality applications.

The funding for the research is provided by the startup fund from the University of Washington, Seattle.